\documentclass[fleqn,12pt]{SelfArx} 

\usepackage[english]{babel} 
\usepackage{tikz}
\usepackage{natbib}
\usepackage{amsmath, amstext}
\usepackage{changepage}
{\setlength{\mathindent}{4mm}

\setlength{\columnsep}{0.55cm} 
\setlength{\fboxrule}{0.75pt} 

\definecolor{color1}{RGB}{0,0,90} 
\definecolor{color2}{RGB}{0,20,20} 

\usepackage{hyperref} 
\urlstyle{same}
\hypersetup{hidelinks,colorlinks,breaklinks=true,urlcolor=color2,citecolor=color1,linkcolor=color1,bookmarksopen=false,pdftitle={Title},pdfauthor={Author}}

\JournalInfo{Robotic Telescopes, Student Research and Education (RTSRE) Proceedings} 
\Archive{ Conference Proceedings, San Diego, California, USA, Jun 18-21, 2017 \\Fitzgerald, M., James, C.R., Buxner, S., Eds. Vol. 1, No. 1, (2018) \\ ISSN XXXX-XXXX (online) / doi : doidoidoidoi / CC BY-NC-ND license \\ Peer Reviewed Article. rtsre.org/ojs} 
\PaperTitle{A Robotic Telescope For University-Level Distance Teaching} 

\Authors{Kolb U.\textsuperscript{1}*, Brodeur M.\textsuperscript{1}, Braithwaite NStJ.\textsuperscript{1}, and Minocha S.\textsuperscript{2}} 
\affiliation{\textsuperscript{1}\textit{School of Physical Sciences, The Open University, Walton Hall, Milton Keynes MK7 6AA, UK}} 
\affiliation{\textsuperscript{2}\textit{School of Computing \& Communications, The Open University, Walton Hall, Milton Keynes MK7 6AA, UK}} 
\affiliation{*\textbf{Corresponding author}: ulrich.kolb@open.ac.uk} 

\newcommand{\keywordname}{Keywords} 

\Abstract{We present aspects of the deployment of a remotely operable telescope for teaching practical science to distance learning undergraduate students. We briefly describe the technical realization of the facility, PIRATE, in Mallorca and elaborate on how it is embedded in the Open University curriculum. The PIRATE teaching activities were studied as part of a wider research project into the importance of realism, sociability and meta-functionality for the effectiveness of virtual and remote laboratories in teaching practical science. We find that students accept virtual experiments (e.g. a telescope simulator) when they deliver genuine, "messy" data, clarify how they differ from a realistic portrayal, and are flagged as training tools. A robotic telescope is accepted in place of on-site practical work when realistic activities are included, the internet connection is stable, and when there is at least one live video feed. The robotic telescope activity should include group work and facilitate social modes of learning. Virtual experiments, though normally considered as asynchronous tools, should also include social interaction. To improve student engagement and learning outcomes a greater situational awareness for the robotic telescope setting should be devised. We conclude this report with a short account of the current status of PIRATE after its relocation from Mallorca to Tenerife, and its integration into the OpenScience Observatories.}

\begin{document}

\maketitle

\begin{tikzpicture}[remember picture,overlay]
   \node[anchor=north west,inner sep=10pt, xshift=1.5cm, yshift=-0.25cm] at (current page.north west)
              {\includegraphics[scale=2.0]{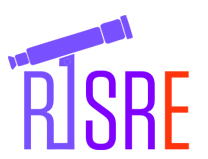}};
\end{tikzpicture}

\flushbottom 

\maketitle 


\thispagestyle{empty} 

\section*{Introduction}

\addcontentsline{toc}{section}{Introduction}

\begin{flushleft}

The Open University has developed its own small-aperture robotic telescope facilities to cater for Open University (OU) distance learning undergraduate and postgraduate students. The OU is the UK's largest provider of part-time distance learning qualifications and derives its name from the practice of open access to any module regardless of prior education. Clear advice and guidance on the hierarchical nature of the Science disciplines, particularly in the Physical Sciences, and on recommended study pathways is given, but there is no formal barrier to enrollment. The student cohort is diverse in background and age, with a median age between 30 and 40 years, and most students holding down a full-time or part-time job. Studying individual modules as a one-off is possible, but students aspiring to a degree in the astronomy domain can opt for a BSc (Hons) entitled "Natural Sciences (Astronomy and Planetary Sciences)" which includes online practical activities at Stages 2 and 3 (OU Stages 2 and 3 are roughly equivalent to years 2 and 3 at a traditional university with full-time students). The strict scheduling needs of time-limited observing activities within the curriculum for student cohorts of order 100, and the specific need to cater for OU students, made it desirable to establish an OU-owned and administered facility rather than buying in on existing robotic telescope networks which come with their own set of constraints and limitations. \\
\bigskip

To this end we initiated a pilot project in 2008, the Physics Innovation Robotic Telescope Explorer (PIRATE), funded by the then Physics Innovation Centre of Excellence in Teaching and Learning (piCETL), one of a number of CETLs created by the Higher Education Funding Council in England (HEFCE). The initial configuration (\cite{lucas2011software}; \cite{kolb2010pirate}), a robotic German mount by Software Bisque, in a simple roll-off shed on the roof of the catering building of the Observatori Astronomic de Mallorca (OAM; N $39 ^\circ$ $38 ^\prime$  $34.31 ^{\prime\prime}$, E $2^\circ$ $57^{\prime}$ $3.34 ^{\prime\prime}$, 160m above sea level), as seen in Figure ~\ref{fig1} (left). The OAM is a teaching observatory that hosted, at the time, several week-long courses in observational astronomy per year for those OU students who were able to travel to Mallorca.  \\
\bigskip

PIRATE underwent upgrades over the years as described in \cite{kolb2014pirate} (see also, \cite{holmes2011pirate}), moving into a robotic 3.5m clam-shell Baader Planetarium dome at the top of the main observatory tower and replacing the Celestron-14 with a PlaneWave CDK17 17 inch astrograph (Figure ~\ref{fig1}, right). The study of the educational use of PIRATE which forms the core of this report relates to this consolidated Mallorca phase of the facility. \\

\begin{figure*}[ht!]
\begin{minipage}[c]{0.45\textwidth}
\includegraphics[width=\textwidth, angle=270]{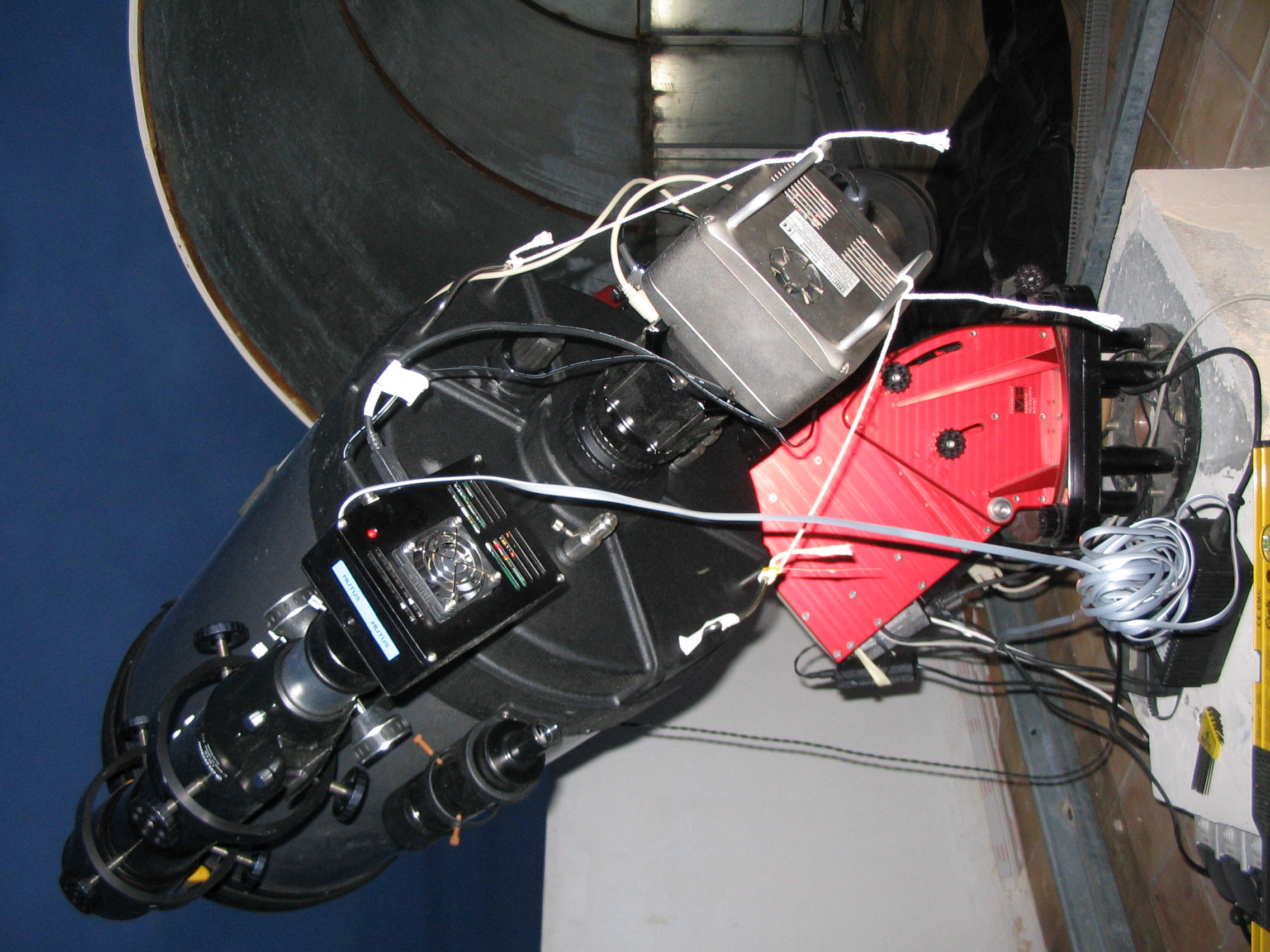}
\end{minipage}
\hfill
\begin{minipage}[c]{0.53\textwidth}
\includegraphics[width=\textwidth]{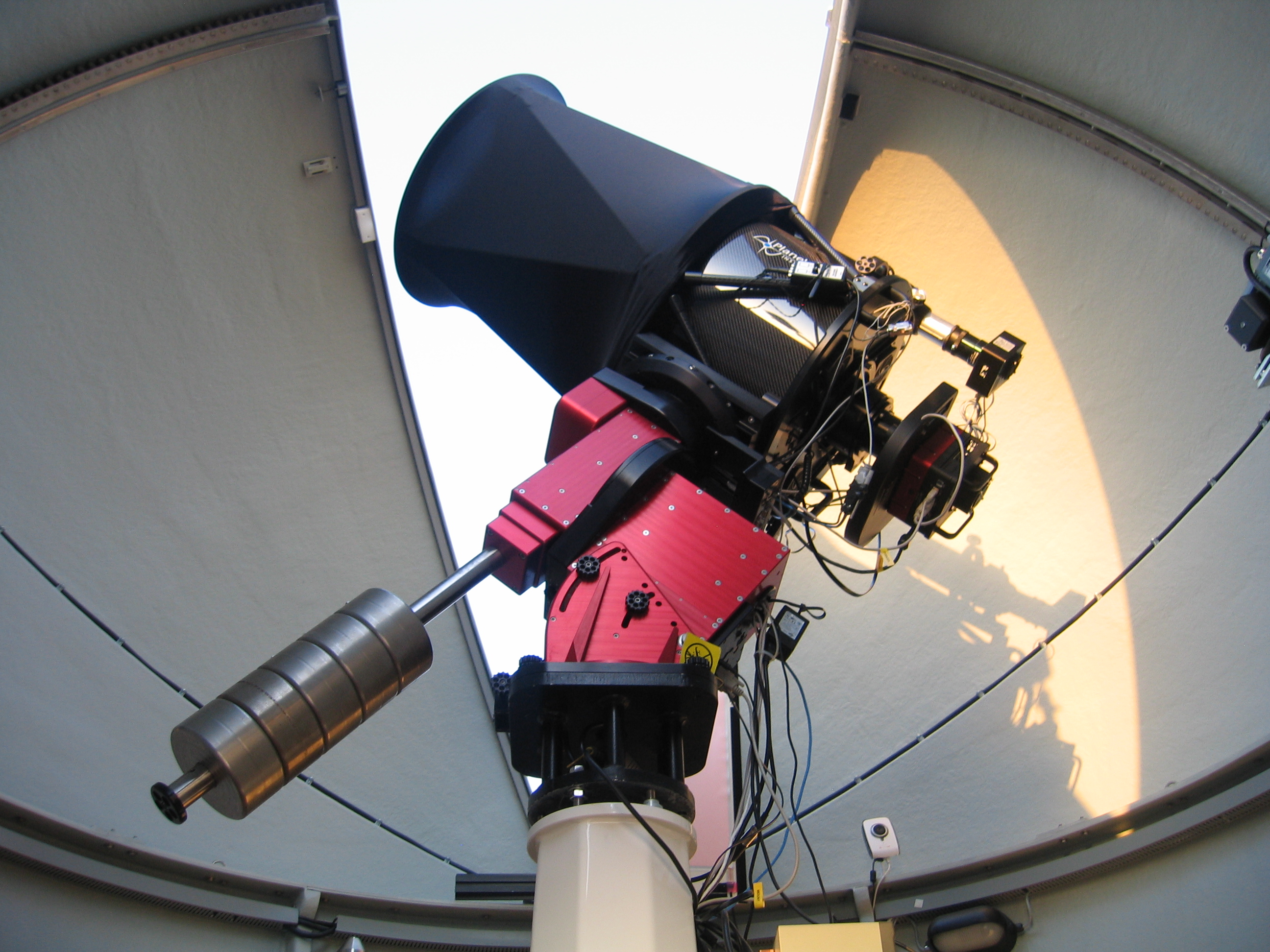}
\end{minipage}%
\caption{\small{\textit{PIRATE at the Observatori Astronomic de Mallorca. Left: Early pilot installation. Right: Consolidated installation in 3.5m dome. Photos by Juan Rodriguez.}}}
\label{fig1}
\end{figure*}

\end{flushleft}

\section*{Curriculum Use}

\addcontentsline{toc}{section}{Curriculum Use}
\begin{flushleft}

There are currently two modules in the OU's curriculum where optical robotic telescopes are deployed to support the teaching of practical science in general, and techniques in observational astronomy in particular. \\
\bigskip

The 30 credit Stage 2 module SXPA288 (previously also SXP288) \textit{Practical science: physics and astronomy} comprises four different 6-7 week long topics in the physical sciences, each with their distinct on-line or virtual practical activities. For context, full-time students would complete 120 credits per year, while most OU students study at a rate of 60 credits per year. The astronomy topic in SXPA288/SXP288 gives the students the choice between two projects. The first project deploys the OU's remotely operable 3m radio dish ARROW (A Robotic Radio telescope Over the Web) to map the distribution of neutral hydrogen in the Galaxy. The second project, which is the focus of the subsequent discussion on how the learning activity was perceived by the students, constructs color magnitude diagrams of star clusters, using PIRATE, to study the cluster distances and ages. In both projects the primary aim is the learning of basic experimental (observational) skills including planning and conducting an experiment, record keeping, awareness of the role of uncertainties, and report writing. This is facilitated by the undertaking of a real scientific observational investigation. The secondary aim is to teach basic methods and techniques in observational astronomy, such as, for the PIRATE strand, CCD image data reduction and aperture photometry. Whilst the data analysis part of either project can be achieved with archival data alone, the center piece of each strand is a series of online session where the students are remotely controlling the hardware in real-time to acquire their own data for later analysis. The practical activity thus conveys the challenges of, in the case of PIRATE, night-sky imaging; it provides ownership of the process and hence a powerful motivation for learners. The 4 hr long PIRATE observing sessions are shared between an observer team of 4 or 5 students, both to increase the number of students with access to the telescope, and to share the workload and responsibilities during the session by peer support. The assessment of the activity, a short write-up presenting and interpreting the colour magnitude diagrams of three star clusters, is prepared by each student individually and contributes to the overall module grade roughly in proportion to the time spent on the activity. \\ 
\bigskip

The 30 credit Stage 3 module S382 \textit{Astrophysics} includes a 10 week long astrophysical data analysis project making up a third of the module. As for SXPA288 the students have a choice between two project flavors, the PIRATE strand and an archival data strand. The former involves the acquisition of photometric data of a periodic variable star coincident with an X-ray source, to classify or help improve the classification of the source, while the latter makes use of the Sloan Digital Sky Survey to construct and interpret a composite spectrum of a quasar. The S382 project is structured into weekly activities and involves groups of about 10 students who collaborate to achieve the project goal. Collaborative working is promoted by activities that explicitly require dividing tasks up between group members, and coordination within the group. To this end in each study week one group member takes on the role of project manager. The outcome of the project is a 3000 word collaborative project report, in the style of a scientific paper, written using a wiki. The communication during the project is mediated by synchronous meetings, mostly via Skype or Blackboard Collaborate, and by asynchronous forums and a wiki. The PIRATE project has three phases. The three week long induction phase culminates in a target selection exercise where the group compiles a shortlist of suitable sources from a catalog of periodic variable stars identified by SuperWASP that are coincident with a ROSAT source \citep{norton2007new}. The academic module lead considers all individual shortlists and approves the actual targets, often doubling up between groups so that in case of bad weather data acquired by a different group can supplement a group's own data. The 4 or 5 week long data acquisition phase typically sees 1 full night of observing per week for each of the groups. Literature research and data analysis is advanced in parallel. The review phase during the last 2-3 weeks of the project is devoted to writing up and editing the final wiki report. The grade on the project is determined from two components, with equal weight. The first component is the project report; this is common to all group members, but down-weighted for those (rare) cases of students who have not satisfactorily engaged with the collaborative aspects of the project. The second assessment component is a portfolio of weekly progress reports, prepared by each students separately, that provide specific evidence of how the student achieved the project module learning outcomes.

\end{flushleft}

\section*{Observing Sessions}

\addcontentsline{toc}{section}{Observing Sessions}

\begin{flushleft}

In both curriculum uses of PIRATE the observing sessions are set up as an interactive experience with real-time, online control, allowing a team of 2-5 student observers to conduct the session's observing program while also linked to each other via voice-only audio. \\
\bigskip

The control system adopted for PIRATE in Mallorca was based on a Windows PC running the observatory control software ACP by DC-3 Dreams, with relevant auxiliary software, including MaxIm DL by Diffraction Ltd to control the camera, FocusMax to administer the auto-focus procedure, and The Sky X as the driver for the mount. The ACP web interface served as the student observer team's main communication portal with the telescope, displaying in real-time status information and thumbnail previews of the latest images, while allowing the user to submit commands and to download their raw data frames as FITS files. All observer team members have simultaneous access to the control interface. Further external tools, not linked into ACP, a webcam video and audio stream, a periodically updating all-sky camera and weather information from a Boltwood cloud sensor completed the diagnostic information available to the observer team. \\
\bigskip

The S382 PIRATE project students also make use of a "virtual telescope", in the form of the \textit{PIRATE simulator}, where the control software ACP is set up in its simulator mode. The user interacts with the simulator in same way via the ACP web interface as the user of the real telescope. The simulator pretends to go through the motions of slewing the telescope, focusing, tracking, image acquisition, guiding, image download etc., therefore no hardware is actually involved. The image returned to the user is a mocked-up fits file of a star field on the basis of a star catalog and an assumed point spread function. The PIRATE simulator was set up behind a booking system offering 2 hour long training slots. Each student was expected to work through at least one training slot in good time before their real observing session, to familiarize themselves with the control interface and the routine activities they will encounter during the observing run. \\
\bigskip

At the beginning of the real-time observing session the members of the observer team are assigned to specific roles, such as telescope operator, log keeper and data analyst, to help organize their collaboration. Meeting fellow students online and speaking to them via the computer has been a novel experience for many SXPA288 and S382 students. The traditional OU distance learner at the time, and in many cases still now, would have been used to exchanging messages on a forum, emailing and perhaps speaking to a tutor on the phone, and more recently attending module-wide online lectures or tutorials using the virtual classroom software Blackboard Collaborate. But even those synchronous events allow reluctant participants to take a backseat, attending but with limited or no interaction, and a majority often watch the recording rather than take part in the live event, either out of choice or because they were unavailable. The real-time observing sessions thus uniquely foster team-building and synchronous collaborative working of distance learners. Weather permitting the S382 sessions ran from dusk to the early hours, for as long as a team was willing to persist. Anecdotal evidence suggests that once the observing program was well underway conversations on science and studying in general progressed to a deeper philosophical level, or to more light-hearted exchanges, and some teams developed a friendship which outlasted the module.\\
\bigskip

When an observer team begins a real-time session for the first time an astronomy tutor with a higher level of access to the remote telescope system, known to the students as the \textit{Night Duty Astronomer} (NDA), welcomes the team on Skype, provides a basic briefing on safe observing, and verifies that the team have a reasonable grasp of what is expected from them. It is also the NDA's responsibility to ensure that the observatory is powered on and fully connected, or to arrange for technical help if that is not the case. The student guidance notes for the observing session are self-contained and written for distance learners who study independently. In principle there should therefore be no need to provide tutor support throughout the session, so the NDA is meant to withdraw after the initial briefing, but has to remain on call in case of technical problems, or any other issues that would prevent the successful conduct of the session. In practice, the NDA often has been playing a more active role, for two main reasons. Perhaps not unexpectedly some observer teams turned up with insufficient time spent on preparing for the session, and some regarded the NDA as a convenient demonstrator who knew all the answers. In such cases the NDA sought to find a balance between helping students along towards acquiring useful data for the project and giving pointers that would allow the team to achieve the learning outcomes in the way intended, by team work using the teaching resource provided. The second reason for occasionally more active NDA involvement was due to the system set-up of PIRATE in Mallorca. The OAM is not connected to the electricity grid and obtains power from an on-site generator. At the time S382 and SXPA288 were presented from Mallorca neither the power supply nor the internet connection speed was perfectly stable, giving rise to occasional system glitches where the telescope would suddenly appear to go offline, or would refuse to execute remote commands. The generally high humidity at the low altitude Mediterranean location increased the hardware's vulnerability to the occasional loss of connection or loss of communication with its control computer. The vast majority of these issues could be swiftly resolved by the NDA using a remote-desktop style connection that allows reconnecting components or rebooting software or even the whole system. The observer teams by and large did not mind as long as the NDA was swiftly at the task, but the overall observing experience was clearly affected by these events. Equally important, the NDA post is a limited resource and not paid as a full-time support astronomer, and nor would the distance teaching of observational astronomy in the OU context be affordable if such a full-time post were required. A sustainable use of remote telescopes in the way the OU curriculum demands requires a robust site infrastructure. The OAM is awaiting investment to achieve this, but recent developments, as described in Section 5 ensured that PIRATE is now on such a robust footing.\\
\bigskip

The rationale for asking observer teams to stay online with PIRATE throughout the session, even into the early hours and when only one target is being monitored in a long time series, was that this is the most effective way to ensure that the data quality remains good throughout the run, that observations can safely resume after intermittent clouds, and that the a technical malfunction can acted upon quickly. At the same time the real-time element fosters team building and enhances collaboration, thus creating a highly motivational learning environment quite unlike what OU distance learners would have been used to. The observer team setting is also accessible to solitary learners, less outgoing students or students who prefer to limit their social interactions, as it is possible to contribute to and benefit from those session by text chat.\\

\end{flushleft}

\section*{Evaluation}

\addcontentsline{toc}{section}{Evaluation}

\begin{flushleft}

The curriculum deployment of PIRATE in Mallorca was evaluated over the period 2013-2015 as part of a PhD research project at the Open University (e.g. \cite{brodeur2016design}; see also \cite{brodeur2014teaching} and \cite{brodeur2014teaching2}). Specifically, the research considered the importance of authenticity, sociability and meta-functionality for the effectiveness of virtual experiments (such as a telescope simulator) and remote-access experiments (such as a robotic telescope) in teaching practical science to undergraduate students. This contributes to a wider critical discussion on the educational merits of remote labs, where some professional researchers and educators have cast doubt on the effectiveness of remote observatories in teaching observational astronomy (e.g. \cite{privon2009importance}; \cite{jacobi2008effect}), while others found that remote experiments can be as effective as proximal ones (e.g. \cite{corter2004remote}; \cite{lowe2013evaluation}).\\
\bigskip

Our study deployed a mixed-methods approach \citep{greene1997defining} on a sample of Open University students, and for the focus on the astronomy aspect included 40 students at Stage 2 and 160 students at Stage 3. The quantitative data comprised anonymized demographic information and assessment scores from student records, as well as pre- and post-activity electronic surveys with ranking questions and Likert-style queries. Qualitative data were obtained from the open-ended survey questions and in semi-structured individual or group interviews conducted via Skype (nine at Stage 2, eight at Stage 3). A detailed account of this study will be published elsewhere (Brodeur et al, in preparation). Here we report only some of the results pertinent to the use of robotic telescopes.\\

\end{flushleft}


\subsection*{Quantitative Findings}

\addcontentsline{toc}{section}{Quantitative Findings}

\begin{flushleft}

The analysis of the survey responses does not reveal any stand-out, strong correlations between student demographics and perceived experience, but there are two notable trends. The first one is that with increasing age students display an increased agreement with the sentiments that both virtual experiments (VEs) and remote experiments (REs) make practical science enjoyable. By contrast younger students expressed a preference for being co-located with practical equipment. This may surprise at first glance as the younger generation has been growing up in an environment where digital devices are ubiquitous and accessing the virtual, online world is an integral part of the daily routine. The older generation have necessarily embraced online technology later in life. Yet younger students may find real labs more desirable just because they are different from the virtual world which has become too common-place. Older students may value remote labs more because they are more easily accessible, more comfortable to use, at a time and place that is convenient in the busy life they are familiar with.\\
\bigskip

The second trend seen in the surveys relates to the assessment outcome, i.e. the grades achieved on the module with the embedded telescope activity. Students who agree with the notion that VEs make practical work more enjoyable tend to obtain higher grades. It is difficult to disentangle if this trend is simply a consequence of the fact that more motivated students are doing better overall, or if indeed those more engaged students who take time to prepare the observing session on the simulator are better prepared for the real observing session to such an extent that they ultimately more readily achieve the overall learning outcomes. Mirroring this, students who replied that REs are not effective for teaching collaboration tend to obtain lower grades on the module. As the telescope activities have a significant on-line collaborative element it is likely that the affected students represent learners who either feel that they are being negatively impacted by fellow students in their group, either because they perceive them as not contributing sufficiently, or because there are students who dominate the group or who are too far ahead and confidence and motivation - or these are students who are genuinely solitary learners who are unduly challenged by the need for group working.\\ 

\end{flushleft}


\subsection*{Qualitative findings}

\addcontentsline{toc}{section}{Qualitative Findings}

\begin{flushleft}

Analysis of the free text survey responses and of interview transcripts reveals insights into the perceived importance of realism, sociability and meta-functionality for PIRATE and its simulator.\\
\bigskip

The ACP web interface and the webcam stream was seen as "realistic", sufficiently so that there was no desire to have access to a virtual reproduction or animation of hardware control panels that some students might be familiar with in the context of an amateur telescope, such as a handset as it is used to drive the mount. Yet the single webcam trained onto the telescope that only shows darkness when the IR beam is turned off for the actual image acquisition, is not enough to convey a sense of being "connected" to the remote telescope. The students also enjoy and demand realistic data and scenarios, as opposed to merely re-tracing idealized steps that deliver perfect versions of the desired measurements or astronomical images. A student said "You learn more when things go wrong", and this implies the desire to be exposed to real-world, messy data. Realism is important, but data-realism, not photorealism.\\
\bigskip

The students valued the provisions of opportunities for social learning. The Skype-mediated voice-only meetings were not seen to be as real or as collaborative as an observer team that is co-located at an observatory, but these synchronous remote meetings provide clearly a very different experience from a synchronous forum, the normal collaborative tool for OU students. Including talking heads in the online meeting was not seen as a potential improvement because the students felt that there were already enough visual clues to deal with during an observing session, including the ACP web interface, the webcam, the weather diagnostics, the observer's log, the fits file viewer and instructions on the Virtual learning Environment or in PDF documents. A student on the autism spectrum also stated that voice-only sessions are preferable to video links. Various suggestions for improving the collaborative learning were made, such as the team should meet and form before the observing session, and the simulator session should be offered as a collaborative activity just like the real observing run, rather than as a session for a single user. \\
\bigskip

A meta-functional element included in the simulator is the accelerated completion of processes that take considerable time in the real world, such as long image exposures or the focus run. Most students appeared to appreciate that that this saves time, but they also felt that such meta-functionality needs to be clearly signposted so as to manage expectations for the real run. Some students would have preferred a toggle to pro-actively choose between "real" behavior and "accelerated" mode.\\

\end{flushleft}


\section*{Summary and Recommendations}

\addcontentsline{toc}{section}{Summary and Recommendations}

\begin{flushleft}

Students accept virtual experiments (telescope simulator) for "hands on" activities when VEs approximate the interfaces used to operate real-world instruments, when VEs deliver genuine, "messy" data, and clarify how they differ from a realistic portrayal, and are flagged as training tools.\\
\bigskip

Students accept remote experiments (the robotic telescope) in place of on-site practical work when REs incorporate realistic activities, a stable internet connection, and at least one live video feed. They should include group work and facilitate social modes of learning.\\
\bigskip

Several recommendations for improving student engagement and learning outcomes from virtual and remote experiments emerge from our study: It is preferable to devise greater situational awareness rather than rely on video conferencing to mimic co-located practical work. The experiments should enable in-person social interactions, and these should be facilitated to form before the actual remote experiment activity. Virtual experiments, though normally considered as asynchronous tools, should also include social interaction.\\

\end{flushleft}


\section*{OpenScience Observatories at Tenerife}

\addcontentsline{toc}{section}{OpenScience Observatories at Tenerife}

\begin{flushleft}

\begin{figure*}[ht]\centering
\includegraphics[width=150mm]{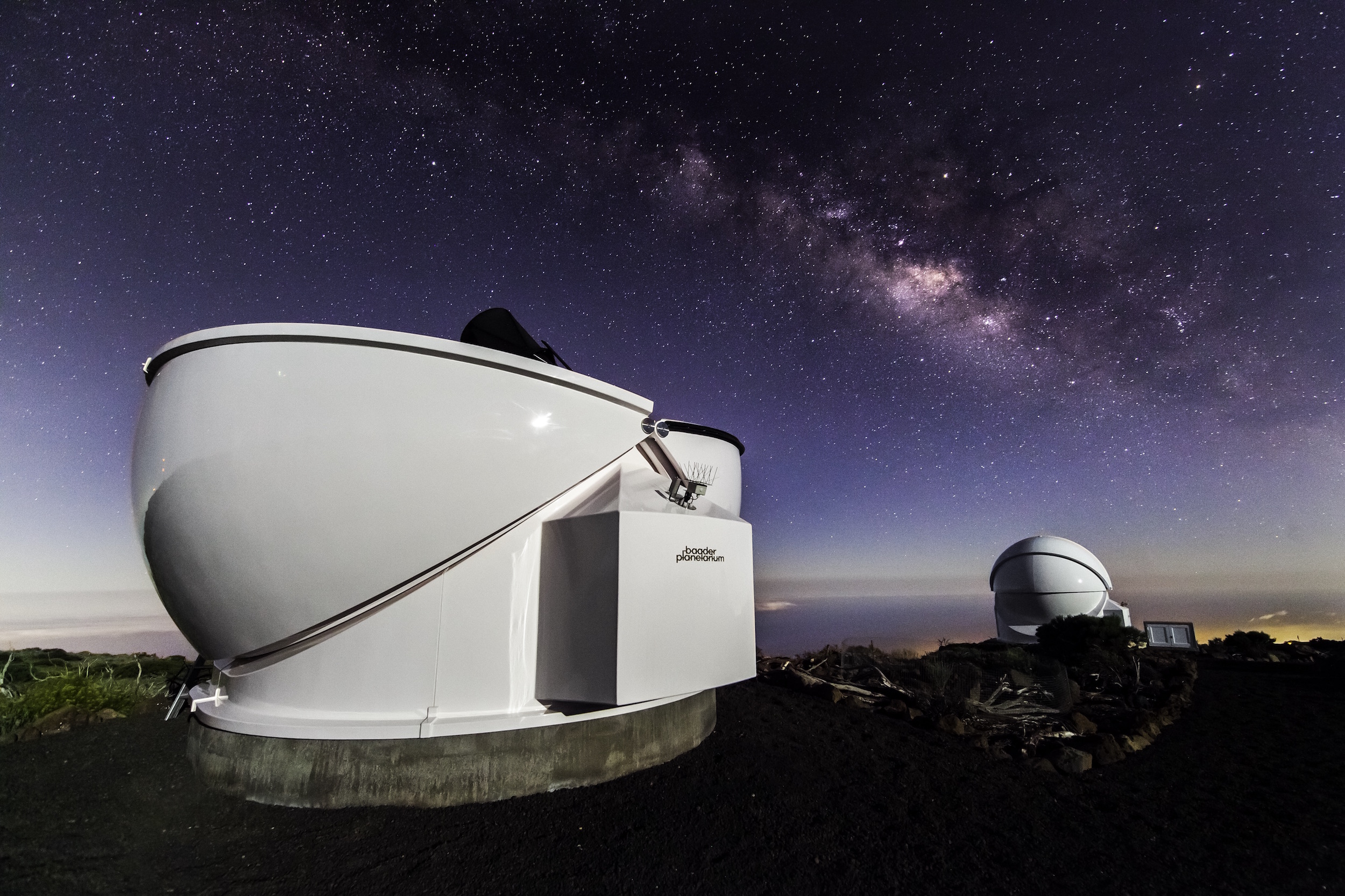}
\captionsetup{format=hang}
\caption{\small{\textit{PIRATE (with 4.5m dome, on the left) and COAST (with 3.5m dome, on the right) at the Observatorio del Teide. Photo by Elena Mora (IAC)}}}
\label{fig:Figure2}
\end{figure*}

We end this account with a brief summary of the latest developments of the PIRATE project that commenced after the end of the curriculum use period considered above. Our evaluation highlighted the importance of the resilience of the hardware set-up and the stability of the internet connection for providing a satisfactory student experience. Technical glitches may deliver some of the desired real-world, messy data and may represent a good learning opportunity, but if pushed too far, or if encountered too frequently, any potential benefit is lost by instilling confusion and denting student confidence.\\
\bigskip

An opportunity to address this issue for PIRATE arose in 2015, when a HEFCE-funded capital infrastructure award enabled a new Open University taught postgraduate qualification, the MSc Space Science and Technology. This opened the door for a step-change that has now completely transformed the PIRATE project and established unprecedented opportunities for teaching and research in observational astronomy for OU students at all Stages. The funding for moving PIRATE to a prime observing site became available in 2015/6, at about the same time when discussions began to transfer the Bradford Robotic Telescope (BRT) \citep{baruch2015robotic} to the OU. Negotiations with the Instituto de Astrof{\'i}sica de Canarias (Spain) led to the relocation of an upgraded PIRATE to the Observatorio del Teide (Tenerife), and to the construction of a second, new facility, COAST (COmpletey Autonomous Service Telescope) to replace the aging hardware of the BRT, and to handle observing requests from the public portal telescope.org. \\
\bigskip

PIRATE is now in a Baader Planetarium 4.5m All-Sky dome, a robotic dome in clam-shell design (Figure~\ref{fig:Figure2}). The optical tube assembly is a 17 inch PlaneWave CDK-17 corrected Dall-Kirkheim astrograph with focal ratio f/6.8, mounted on 10Micron's German equatorial mount GM4000. The FLI ProLine KAF-16803 imagining camera has $4096^2$ 9 micron pixels, giving a $43 ^\prime$ field of view and $0.63^{\prime\prime}$/px plate scale. The camera is equipped with Baader LRGB broadband filters and H$\alpha$, SII and OIII narrowband filters. The control software ABOT by Sybilla Technology, a much developed version of what is described in \cite{sybilski2014software}, provides both a web control interface for real-time use and manages the fully autonomous queue-scheduled mode of operation. The public front-end telescope.org inherited from the BRT project is passing scheduled requests to Abot for execution. \\
\bigskip

The second facility, COAST, resides next to PIRATE in a smaller Baader Planetarium 3.5m All-Sky dome. A 14 inch Celestron-14 Schmidt Cassegrain telescope with focal ratio f/11 is mounted on a GM4000. The legacy SBIG STL-1001E imaging camera has $1024^2$ 24 $\mu$ pixels, giving a $29^\prime$ field of view and $1.7^{\prime\prime}$/px plate scale. The camera is equipped with Baader Johnson-Cousin BVR broadband filters and H$\alpha$, SII and OIII narrow-band filters. The control software is also ABOT. \\
\bigskip

The large number of usable dark hours per year makes the facilities ideally suited for the highly constrained time-tabled use required for the wider OU curriculum. An assessment of the effectiveness of our new facilities and the ABOT control interface in teaching observational techniques will be the subject of a forthcoming study. \\
\bigskip

PIRATE and COAST form now the Tenerife facilities of the OpenScience Observatories, the Astronomy "wing" of the OpenSTEM Labs, an award-winning major Open University initiative to bring practical science to distance learners world-wide.\\

\end{flushleft}


\section*{Acknowledgements}

\addcontentsline{toc}{section}{Acknowledgements}

\begin{flushleft}

The PIRATE project is a team effort and would not have been feasible without the many contributions by current and former members of the PIRATE crew - too numerous to list here (see \href{http://pirate.open.ac.uk/contact.html}{http://pirate.open.ac.uk/contact.html} and \href{http://www.telescope.org/contact-us.php}{http://www.telescope.org/contact-us.php}). We are grateful to the staff and volunteers of the Observatori Astronomic de Mallorca who helped maintain the facility during its Mallorca period and who provided local emergency assistance. Baader Planetarium played a leading role in the planning and construction of the Tenerife facilities. We thank their staff for their continuous support of the project, and in particular Ladislav Rehak who spent countless hours keeping PIRATE afloat for its curriculum use while it was temporarily based in Mammendorf. We are indebted to John Baruch who made the move of the pioneering telescope.org project to the OU possible. Don Pollacco and John Baruch planted the seeds for the recent relocation of PIRATE to Tenerife. We thank Piotr Sybilski and his team at Sybilla Technologies for their collaboration in moving the project forward. Finally we acknowledge again funding by the HEFCE for the OpenSTEM Labs initiative, and by the Wolfson Foundation.

\end{flushleft}


\bibliographystyle{apalike}

\bibliography{references}

\begin{thebibliography}{}

\bibitem[Baruch, 2015]{baruch2015robotic}
Baruch, J. (2015).
\newblock A robotic telescope for science and education.
\newblock {\em Astronomy \& Geophysics}, 56(2):2.18--2.21.

\bibitem[Brodeur et~al., 2014a]{brodeur2014teaching}
Brodeur, M., Kolb, U., Minocha, S., and Braithwaite, N. (2014a).
\newblock Teaching undergraduate astrophysics with {PIRATE}.
\newblock {\em Revista Mexicana de Astronom{\'\i}a y Astrof{\'\i}sica},
  45:129--132.

\bibitem[Brodeur et~al., 2014b]{brodeur2014teaching2}
Brodeur, M., Kolb, U., Minocha, S., and Braithwaite, N. (2014b).
\newblock Teaching undergraduate astrophysics with {PIRATE}.
\newblock {\em Revista Mexicana de Astronom{\'\i}a y Astrof{\'\i}sica},
  45:133--134.

\bibitem[Brodeur, 2016]{brodeur2016design}
Brodeur, M.~S. (2016).
\newblock {\em Design priorities for online laboratories in undergraduate
  practical science}.
\newblock PhD thesis, Open University.

\bibitem[Corter et~al., 2004]{corter2004remote}
Corter, J.~E., Nickerson, J.~V., Esche, S.~K., and Chassapis, C. (2004).
\newblock Remote versus hands-on labs: A comparative study.
\newblock In {\em Frontiers in Education, 2004. FIE 2004. 34th Annual}. IEEE.
\newblock F1G-17-21.

\bibitem[Greene and Caracelli, 1997]{greene1997defining}
Greene, J.~C. and Caracelli, V.~J. (1997).
\newblock Defining and describing the paradigm issue in mixed-method
  evaluation.
\newblock {\em New directions for evaluation}, 1997(74):5--17.

\bibitem[Holmes et~al., 2011]{holmes2011pirate}
Holmes, S., Kolb, U., Haswell, C., Burwitz, V., Lucas, R., Rodriguez, J.,
  Rolfe, S., Rostron, J., and Barker, J. (2011).
\newblock {PIRATE}: a remotely operable telescope facility for research and
  education.
\newblock {\em Publications of the Astronomical Society of the Pacific},
  123(908):1177--1187.

\bibitem[Jacobi et~al., 2008]{jacobi2008effect}
Jacobi, I.~C., Newberg, H.~J., Broder, D., Finn, R.~A., Milano, A.~J., Newberg,
  L.~A., Weatherwax, A.~T., and Whittet, D.~C. (2008).
\newblock Effect of night laboratories on learning objectives for a nonmajor
  astronomy class.
\newblock {\em Astronomy Education Review}, 7:66--73.

\bibitem[Kolb, 2014]{kolb2014pirate}
Kolb, U. (2014).
\newblock The {PIRATE} facility: at the crossroads of research and teaching.
\newblock {\em Revista Mexicana de Astronom{\'\i}a y Astrof{\'\i}sica},
  45:16--19.

\bibitem[Kolb et~al., 2010]{kolb2010pirate}
Kolb, U., Lucas, R., Burwitz, V., Holmes, S., Haswell, C., Rodgriguez, J.,
  Rolfe, S., Rostron, J., and Barker, J. (2010).
\newblock {PIRATE}-the {piCETL} astronomical telescope explorer.
\newblock {\em \textnormal{In: Norton, Andrew ed.,} Electronic Resources for
  Teaching and Learning}.
\newblock Milton Keynes: The Open University, pp 58-64.
  (http://oro.open.ac.uk/26018/).

\bibitem[Lowe et~al., 2013]{lowe2013evaluation}
Lowe, D., Newcombe, P., and Stumpers, B. (2013).
\newblock Evaluation of the use of remote laboratories for secondary school
  science education.
\newblock {\em Research in Science Education}, 43(3):1197--1219.

\bibitem[Lucas and Kolb, 2011]{lucas2011software}
Lucas, R. and Kolb, U. (2011).
\newblock Software architecture for an unattended remotely controlled
  telescope.
\newblock {\em Journal of the British Astronomical Association},
  121(5):265--269.

\bibitem[Norton et~al., 2007]{norton2007new}
Norton, A.~J., Wheatley, P., West, R., Haswell, C., Street, R., Cameron, A.~C.,
  Christian, D., Clarkson, W., Enoch, B., Gallaway, M., et~al. (2007).
\newblock New periodic variable stars coincident with {ROSAT} sources
  discovered using {SuperWASP}.
\newblock {\em Astronomy \& Astrophysics}, 467(2):785--905.

\bibitem[Privon et~al., 2009]{privon2009importance}
Privon, G.~C., Beaton, R.~L., Whelan, D.~G., Yang, A., Johnson, K., and Condon,
  J. (2009).
\newblock The importance of hands-on experience with telescopes for students.
\newblock {\em arXiv preprint arXiv:0903.3447}.

\bibitem[Sybilski et~al., 2014]{sybilski2014software}
Sybilski, P.~W., Paw{\l}aszek, R., Koz{\l}owski, S.~K., Konacki, M., Ratajczak,
  M., and He{\l}miniak, K.~G. (2014).
\newblock Software for autonomous astronomical observatories: challenges and
  opportunities in the age of big data.
\newblock In {\em Software and Cyberinfrastructure for Astronomy III}, volume
  9152, page 91521C. International Society for Optics and Photonics.

\end{thebibliography}

\end{document}